\documentstyle[12pt]{article}
\title{A New Parameterization for the Lagrangian Density of
Relativistic Mean Field Theory}
\author{G.A.  Lalazissis$^{1,2}$, J. K\"onig$^1$ and P. Ring$^1$ \\
$^1$Physik Department, Technische Universit\"at M\"unchen \\
D-85747 Garching, Germany\\
$^2$ Department of Theoretical Physics\\
Aristotle University of Thessaloniki\\
GR-54006 Thessaloniki, Greece}

\begin{document}
\maketitle
\begin{abstract}
A new parameterization for an effective non-linear
Lagrangian density of relativistic mean field (RMF) theory
is proposed, which is able to provide an excellent
description not only for the properties of stable nuclei
but also for those far from the valley of beta-stability.
In addition recently measured superdeformed mimima in
the Hg-region are reproduced with high accuracy.  
\end{abstract}

Relativistic Mean Field (RMF) \cite{SW.86} theory has
recently gained considerable success in describing various
facets of nuclear structure properties. With a very limited
number of parameters, RMF theory is able to give a quantitative
description of ground state properties of spherical and
deformed nuclei \cite{GRT.90,Rin.96} at and away from the stability
line.  Recently it has been shown that RMF theory is
successful in reproducing the anomalous kink in the isotope
shifts of Pb nuclei \cite{SLR.93} and a first-ever
microscopic description of anomalous isotopic shifts in Sr
and Kr chains \cite{LS.95} has been provided.  Such an
anomalous behavior is a generic feature of deformed nuclei
which include almost all isotopic chains in the rare-earth
region \cite{LSR.96} where RMF theory has been shown to
have a remarkable success.  Moreover good agreement with
experimental data has been found recently for collective
excitations such as giant resonances \cite{VBR.94} and for
twin bands in rotating superdeformed nuclei \cite{KR.93}. 
It is also noted that cranked RMF theory provides an excellent
description of of superdeformed rotational bands in the
A=140-150 region \cite{AKR.96a}, in the Sr region
\cite{AKR.96} and in the Hg region \cite{KR.96}

The starting point of RMF theory is a standard Lagrangian
density \cite{GRT.90}
\begin{eqnarray}
{\cal L}&=&\bar\psi\left(\gamma(i\partial-g_\omega\omega
-g_\rho\vec\rho\vec\tau-eA)-m-g_\sigma\sigma
\right)\psi
\nonumber\\
&&+\frac{1}{2}(\partial\sigma)^2-U(\sigma )
-\frac{1}{4}\Omega_{\mu\nu}\Omega^{\mu\nu}
+\frac{1}{2}m^2_\omega\omega^2\nonumber\\
&&-\frac{1}{4}{\vec{\rm R}}_{\mu\nu}{\vec{\rm R}}^{\mu\nu}
+\frac{1}{2}m^2_\rho\vec\rho^{\,2}
-\frac{1}{4}{\rm F}_{\mu\nu}{\rm F}^{\mu\nu}
\end{eqnarray}
which contains nucleons $\psi$ with mass $m$,  $\sigma$-,
$\omega$-, $\rho$-mesons, the electromagnetic field and
non-linear self-interactions of the $\sigma$-field,
\begin{equation}
U(\sigma)~=~\frac{1}{2}m^2_\sigma\sigma^2+\frac{1}{3}g_2\sigma^3+
\frac{1}{4}g_3\sigma^4.
\end{equation}

The Lagrangian parameters are usually obtained by a fitting
procedure to some bulk properties of a set of spherical
nuclei \cite{Rei.89}.  Among the existing parameterizations
the most frequently used are NL1 \cite{RRM.86}, NL-SH
\cite{SNR.93} and the parameter set PL-40 \cite{Rei.88},
which has been proved to provide reasonable fission
barriers. NL1 and NL-SH sets give good results in most
of the cases.  Along the beta stability line NL1 gives
excellent results for binding energies and charge radii, in
addition it provides an excellent description of the
superdeformed bands \cite{AKR.96a,KR.96}. However, in going away
from the stability line the results are less satisfactory.
This can be partly attributed to the large asymmetry
energy $J\simeq$44 MeV predicted by this force. In addition
the calculated neutron skin thickness shows systematic
deviations from the experimental values for the set NL1
\cite{SR.92}. In the parameter set NL-SH this problem
was treated in a better way and improved isovector
properties have been obtained with an asymmetry energy of
$J\simeq$ 36 MeV. Moreover NL-SH seems to describe the
deformation properties in a better way than NL1. However,
the NL-SH parameterization produces a slight over-binding
along the line of beta-stability and in addition it fails
to reproduce successfully the superdeformed minima in
Hg-isotopes in constraint calculations for the energy
landscape. A remarkable difference between the two
parameterizations are the quite different values predicted
for the nuclear matter incompressibility. NL1 predicts a
small value (K=212 MeV) while with NL-SH a very large value
(K=355 MeV) is obtained. Both forces fail to reproduce the
experimental values for the isoscalar giant monopole
resonances for Pb and Zr nuclei. The NL1 parameterization
underestimates the empirical data by about 2 MeV while
NL-SH overestimates it by about 2 MeV.
  
The aim of the present investigation is to provide a new
improved set of Lagrangian parameters, which to some extend
cures the deficiencies of the existing parameterizations.
For this reason a multi-parameter fit was performed in the
the same way as with the other parameterizations
\cite{Rei.89,SNR.93}.  The nucleon mass was fixed to 939
MeV.  The Lagrangian parameters are the meson masses
m$_{\sigma}$, m$_{\omega}$, m$_{\rho}$, the corresponding
coupling constants g$_{\sigma}$, g$_{\omega}$, g$_{\rho}$
and the parameters g$_2$, g$_3$ of the non-linear potential
U($\sigma$). Apart from the mass of the $\rho$ meson which
was fixed to the empirical value (763 MeV) all the others
were taken as free parameters. The nuclear properties
fitted are the charge radii, the binding energies, and the
available neutron radii of several spherical nuclei. The
experimental input for finite nuclei used in the fitting
procedure is shown in Table 1 in parentheses.  We recall
that for the determination of NL-SH parameters six nuclei
were used in the fit, namely $^{16}O$, $^{40}Ca$,
$^{90}Zr$, $^{116}Sn$, $^{124}Sn$ and $^{208}Pb$ while for
NL1 $^{48}Ca$ and $^{58}Ni$ were also taken into account.
It is noted that for NL1 the experimental information used
was the total binding energies, the diffraction radii and
the surface thickness. For NL-SH  charge radii and neutron
radii were used instead of the diffraction radii and the
surface thickness.  In the present work the number of
nuclei used in the fit was increased to ten. In order to
take into account a larger variation in isospin, in
addition to the eight nuclei used for NL1 the doubly closed
shell nucleus $^{132}$Sn as well as the heavier lead
isotope $^{214}$Pb were also included in the fit. The
experimental values for the total binding energies were
taken from the experimental mass tables \cite{AW.93}, the
charge radii from Ref.  \cite{VJV.87}. The available
neutron radii are from Ref. \cite{BFG.89}. In the case
of open shell
nuclei pairing was considered in the BCS formalism. The
gap parameters $\Delta_{n(p)}$ were determined from the
observed odd-even mass differences \cite{AW.93}.
Specifically, for $^{58}Ni$,  $\Delta_{n}$=1.4 MeV, for
$^{90}Zr$  $\Delta_{p}$= 1.12 MeV, for the two $Sn$
isotopes (A=116,124) the  $\Delta_{n}$ values are 1.17 and
1.32 MeV respectively and finally for $^{214}Pb$
$\Delta_{n}$=0.7 MeV.  The binding energies and charge
radii were taken within an accuracy of 0.1\% and 0.2\%
respectively. For the neutron radii, however, due to
existing uncertainties the experimental error taken into
account was 2\%.  In addition in the fitting procedure some
nuclear matter properties were also considered. As
``experimental input'' the following values were used: E/A=
-16.0 MeV (5\%), $\rho$ = 0.153 (fm$^{-3}$ (10\%), K = 250
(MeV) (10\%) J = 33 MeV (10\%). The values in parentheses
correspond to the error-bars used in the fit.

In Table 1 we list the predictions of NL3 for the ground
state properties of the nuclei used in the fit. It is seen
that they are in very good agreement with the empirical values.

In Table 2 we show the values for the new parameter set.
Adopting the convention introduced by P.-G. Reinhard
\cite{Rei.89,RRM.86,Rei.88} for the non-linear
parameterizations the set is named NL3.

In the same table we give nuclear matter properties calculated
with NL3. The saturation density $\rho$ has the value
0.1483 fm$^{-3}$.  The effective mass $m^*/m$ was found
0.6. It is the same as for NL-SH and slightly higher than
for NL1. The nuclear matter incompressibility has the
value $K=$271.8 MeV. It is therefore somewhere in the middle
between the values predicted by NL1 and NL-SH.  Finally the
asymmetry energy J is 37.4 MeV. It is closer to that of 
NL-SH and much smaller than that of NL1.

In the following we present some applications of the new
parameter set NL3 using the various RMF codes of the Munich
group.  We performed detailed calculations for the chain of Sn
isotopes with the spherical Relativistic Hartree Bogoliubov
(RHB) code discussed in Ref. \cite{GEL.96}. In Fig. 1 we
show the isotopic dependence of the deviation of the
theoretical mass calculated in RMF theory from the
experimental values \cite{AW.93} for Sn nuclei. The
theoretical results were obtained using the parameter sets 
NL1, NL-SH and NL3. It is seen that all parameterizations
give a very good description of the experimental masses. It
is also seen, however, that the new force NL3 is able to
provide improved results over the NL1 and NL-SH, reducing
rms deviation of the masses.

Axially symmetric calculations have been performed for some
well deformed rare-earth and actinide nuclei using the new
Lagrangian parameterization NL3. Here the pairing
correlations are taken into account using the BCS
formalism. The pairing parameters $\Delta_{n(p)}$ were
taken from tables XI and XIII of Ref. \cite{GRT.90}.  In
Table 2 we give the results of our calculations together
with the experimental information whenever available. It is
seen that NL3 gives excellent results for the ground
state properties of rare-earth and actinide nuclei.  The
experimental masses \cite{AW.93} are reproduced within an
accuracy of a few hundreths of keV. The charge radii are in
very good agreement with the experiment \cite{VJV.87}. The
deformation properties are also in excellent agreement with
the empirical values. The absolute values of the empirical
$\beta_2$ were obtained from the compilation of Raman et
al. \cite{RMM.87} The experimental data for the hexadecupole
moments of rare-earth nuclei are from a very recent
compilation by L\"obner \cite{Loe.95}. Finally the
experimental data for the proton quadrupole moments were
taken from tables XII and XIV or Ref.. \cite{GRT.90}. Next
we report some preliminary results for the Giant monopole
breathing energies of $^{208}Pb$ and $^{90}Zr$ nuclei
obtained from generator coordinate calculations based
on constrained RMF wave functions. A detailed study
including in addition dynamic RMF calculations will appear 
in a forthcoming publication \cite{Vre.96}. In Table 4 we show
results of calculations using the new parameter set NL3 and
compare it with experimental results and calculations
obtained from the sets NL-SH and NL1.  It is seen that NL3
is able to reproduce nicely the experimental values while
the other two forces fail, either underestimating (NL1) or
overestimating (NL-SH) the experiment by almost 2 MeV. This
is an indication that NL3 has a correct value for the
nuclear incompressibility.

Recently, the excitation energy between the ground state band and 
the superdeformed band in $^{194}$Hg was measured for the
first time \cite{Kho.96}. Extrapolating to zero angular 
momentum the superdeformed minimum was found to be 6 MeV
above the ground state. Performing RMF calculations with 
the parameter set NL3 and mapping the energy surface by
a quadratic constraint we found the superdeformed minimum
at an excitation energy of 5.997 MeV above the ground state.
A detailed study will be published elsewhere \cite{Lal.96}.

In conclusion our calculations with the new RMF parameterization 
NL3 give very good results in all cases considered so far. 
It is in excellent agreement with experimental nuclear masses, 
as well as the deformation properties. For the first time a RMF
parameterization reproduces the isoscalar monopole energies
of Pb and Zr nuclei. This gives us confidence that NL3 can 
be used successfully in future investigations together with
the other parameterizations.
  
One of the authors (G.A.L) acknowledges support by the
European Union under the contract TMR-EU/ERB FMBCICT-950216.
This work is also supported in part by the
Bundesministerium f\"ur Bildung und Forschung under the
project 06 TM 743 (6). We thank Prof. K.E.G. L\"obner for
supplying us with his compilation on hexadecupole moments prior
to its publication.

\newpage
\begin{table}
\caption{ The total binding energies, charge radii and neutron radii used
    in the fit (values in parentheses) together with the NL3 predictions.} 
\begin{center}
\begin{tabular}{llll}
\hline
Nucleus   &~~B.E (MeV) &~~ r$_{ch}$  (fm) &~~ r$_{n}$ (fm) \\   
\hline
   $^{16}$O  &-128.83  (-127.62) &2.730 (2.730) & 2.580         \\ 
   $^{40}$Ca &-342.02  (-342.06) &3.469 (3.450) & 3.328 (3.370) \\
   $^{48}$Ca &-415.15  (-416.00) &3.470 (3.451) & 3.603 (3.625) \\
   $^{58}$Ni &-503.15  (-506.50) &3.740 (3.769) & 3.740 (3.700) \\
   $^{90}$Zr &-782.63  (-783.90) &4.287 (4.258) & 4.306 (4.289) \\
  $^{116}$Sn &-987.67  (-988.69) &4.611 (4.627) & 4.735 (4.692) \\
  $^{124}$Sn &-1050.18 (-1049.97)&4.661 (4.677) & 4.900 (4.851) \\
  $^{132}$Sn &-1105.44 (-1102.90)&4.709         & 4.985         \\
  $^{208}$Pb &-1639.54 (-1636.47)&5.520  (5.503)& 5.741 (5.593) \\
  $^{214}$Pb &-1661.62 (-1663.30)&5.581  (5.558)& 5.855         \\
\hline
\end{tabular}
\end{center}
\end{table}
%
\begin{table}
\caption{Parameters of the effective interaction NL3 in the RMF theory
together with the nuclear matter properties obtained with this effective 
force.}
\begin{center}
\begin{tabular}{ll}
\hline
  $M$ = 939 (MeV)  &        \\
  $m_{\sigma}$ = 508.194 (MeV) & $g_{\sigma}$ = 10.217 \\
  $m_{\omega}$ = 782.501 (MeV) & $g_{\omega}$ = 12.868 \\
  $m_{\rho}$   = 763.000 (MeV) & $g_{\rho}$   =  4.474 \\
  $g_2$       = -10.431 (fm$^{-1}$) & $g_3$       = -28.885 \\ 
\                        \\ 
\hline
Nuclear matter properties \\
\hline
\                          \\
 $\rho_0$       & 0.1483 fm$^{-3}$ \\
 $(E/A)_\infty$ & 16.299 MeV       \\
 $K$            & 271.76 MeV       \\
 $J$            & 37.4   MeV       \\
 $m^*/m$        & 0.60             \\
\end{tabular}
\end{center}
\end{table}
%
\begin{table}
\caption{ Total binding energies B.E (in MeV) , charge radii r$_c$ (in fm),
quadrupole deformation parameters $\beta_{2}$,  proton quadrupole moments 
$q_{p}$ (in barns) and proton hexadecupole ($h_{p}$) moments (in barns$^2$)
for some deformed rare-earth and actinide nuclei with the parameterization
NL3. The values in parentheses correspond to the empirical data. For details
see text}   
\begin{center}
\begin{tabular}{llllll}
\hline
A & B.E & r$_{c}$ &$\beta_{2}$ &Q$_{p}$ & H$_{p}$ \\
\hline
$^{152}Sm$ \\
 &~-1294.49&~5.177&~0.301&~5.63&~0.48\\
 &(-1294.05)&(5.099)&(0.306)&(5.78)&(0.46(2))\\ 
$^{158}Gd$ \\
 &~-1296.40&~5.176&~0.342&~7.14&~0.48\\
 &(-1295.90)&(5.172)&(0.348)&(7.36)& (0.39(9))\\
$^{162}Dy$ \\
  &~-1324.09&~5.227&~0.347&~7.54&~0.45\\
  &(-1324.11)&(5.210)&(0.341)&(7.36)&(0.27(10))\\
$^{166}Er$ \\
     &~-1351.06  &~5.272&~0.349&~7.87&~0.36\\ 
     &(-1351.57)&(5.303)&(0.342)&(7.70)& (0.32(16))\\ 
$^{174}Yb$ \\
     &~-1406.15   &~5.336      &~0.328&~7.77&~0.04\\ 
     &(-1406.60) &(5.410)    &(0.325)&(7.58)&(0.22$^{+0.14}_{-0.18}$)\\
\     \\
$^{232}Th$ \\
     &~-1766.29&~5.825&~0.251&~9.23&~1.06\\
     &(-1766.69)&(5.790)&(0.261)&(9.62)&(1.22)\\
$^{236}U$  \\
     &~-1790.67  &~5.873 &~0.275&~10.60 &~1.16     \\
     &(-1790.42) &       &(0.282)&(10.80) & (1.30) \\
$^{238}U$ \\
    &~-1801.39&~5.892&~0.283&~10.93&~1.07 \\ 
    &(-1801.69)&(5.854)&(0.286)&(11.12)&(1.38)\\
\hline
\end{tabular}
\end{center}
\end{table}
%
\begin{table}
\caption{Isoscalar giant monopole energies calculated with the effective
interactions NL3, NL1, NL-SH along with the empirical values. }
\begin{center}
\begin{tabular}{lllll}
\hline
A         &  expt.          & NL3  & NL1  & NL-SH \\
\hline
$^{208}Pb$&  13.8 $\pm$0.5  & 13.0 & 11.0 & 15.0  \\
$^{90}Zr$ &  16.2 $\pm$0.5  & 16.9 & 14.1 & 19.5  \\
\hline
\end{tabular}
\end{center}
\end{table}
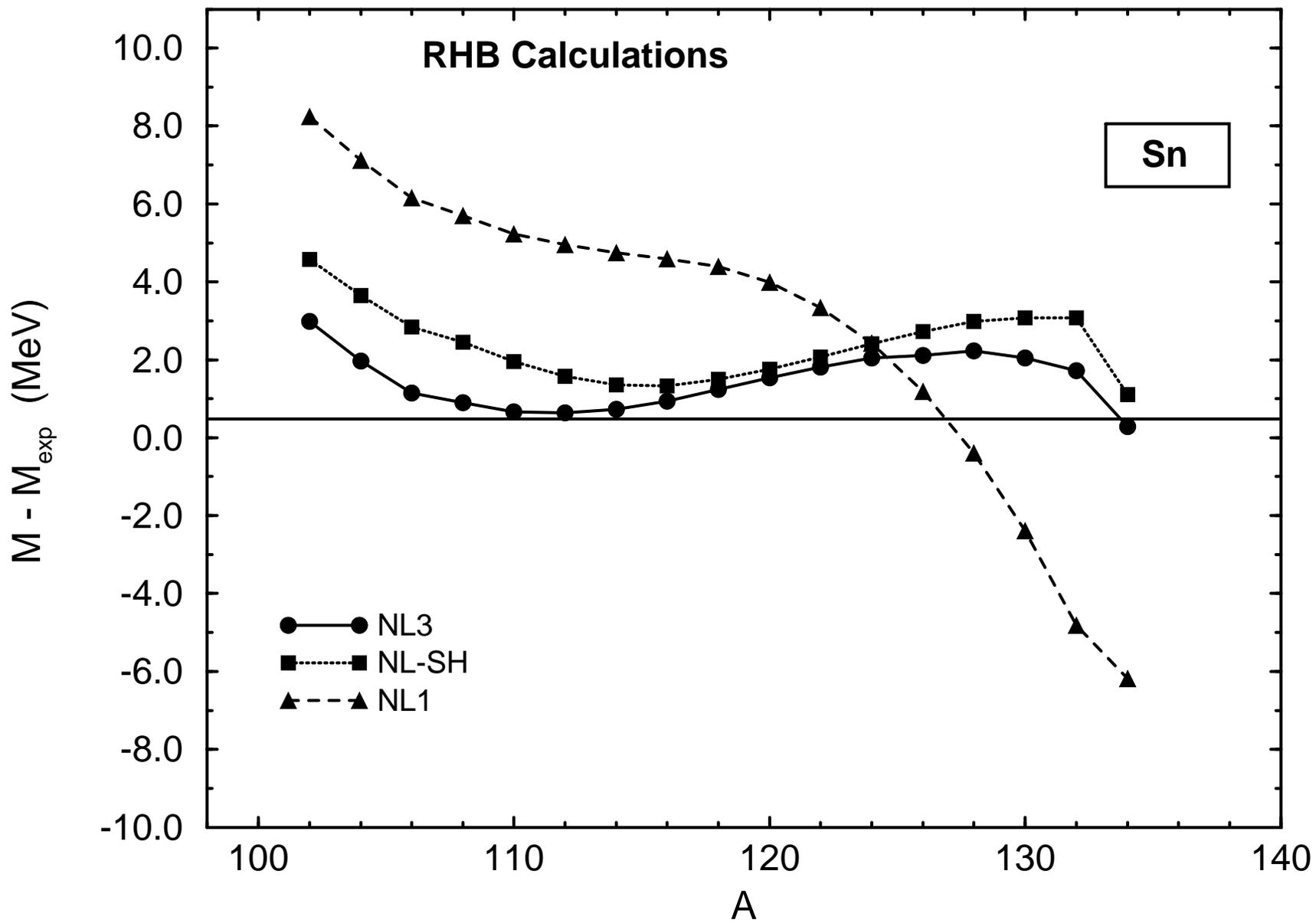
\end{document}